\documentclass[journal]{IEEEtran}
\usepackage[T1]{fontenc}
\usepackage[latin9]{inputenc}
\usepackage{amsmath}
\usepackage{amssymb}
\usepackage{amsthm,bm}
\usepackage{url}
\usepackage{cite}
\usepackage{babel,verbatim,balance}
\usepackage[rgb]{xcolor}
\usepackage{graphics, graphicx}
\usepackage{acronym}
\usepackage{upgreek}
\usepackage[bookmarks,colorlinks]{hyperref}
\usepackage{setspace,gensymb}
\usepackage{booktabs}

\usepackage{float}
\usepackage{pgfplots}
\usepackage[bookmarks,colorlinks]{hyperref}
\usepackage[linesnumbered,ruled,lined]{algorithm2e}
\usepackage{enumitem}
\usepackage{algpseudocode}
\usetikzlibrary{shapes.multipart,intersections}
\usepackage{cite}
\usepackage{amsmath,amssymb,amsfonts,amsthm,steinmetz}
\usepackage{graphicx}
\usepackage{mathrsfs}  
\usepackage{textcomp}
\usepackage{acronym}
\usepackage{xcolor}
\usepackage{upgreek,xspace}

\usepackage{tikz}
\usetikzlibrary{calc}
\makeatletter
\newcommand{\gettikzxy}[3]{%
  \tikz@scan@one@point\pgfutil@firstofone#1\relax
  \edef#2{\the\pgf@x}%
  \edef#3{\the\pgf@y}%
}

\begin{document}

\title{End-to-End Dynamic Metasurface Antenna Wireless System: Prototype, Opportunities, and Challenges}

\author{Fran\c{c}ois Yven, Jean Tapie, J\'er\^ome Sol, and Philipp del Hougne
\thanks{The authors are with Univ Rennes, CNRS, INSA Rennes, IETR - UMR 6164, F-35000, Rennes, France.
}
\thanks{F.~Yven and J.~Tapie contributed equally to this work.}
\thanks{\textit{Corresponding Author: Philipp del Hougne (e-mail: philipp.del-hougne@univ-rennes.fr).}}
\thanks{This work was supported in part by the ANR France 2030 program (project ANR-22-PEFT-0005), the ANR PRCI program (project ANR-22-CE93-0010), the French Defense Innovation Agency (project 2024600), the European Union's European Regional Development Fund, and the French Region of Brittany and Rennes M\'etropole through the contrats de plan \'Etat-R\'egion program (projects ``SOPHIE/STIC \& Ondes'' and ``CyMoCoD'').}
}

\newcommand{\gzero}{\textcolor{gray}{0}}

\maketitle
\begin{abstract}
Dynamic metasurface antennas (DMAs) are a promising hybrid analog/digital beamforming technology to realize next-generation wireless systems with low cost, footprint, and power consumption. 
The research on DMA-empowered wireless systems is still at an early stage, mostly limited to theoretical studies under simplifying assumptions on the one hand and a few antenna-level experiments on the other hand. Substantial knowledge gaps arise from the lack of  complete end-to-end DMA-empowered wireless system prototypes. In addition, recently unveiled benefits of strong inter-element mutual coupling (MC) in DMAs remain untapped.
Here, we demonstrate a K-band prototype of an end-to-end wireless system based on a DMA with strong inter-element MC. 
To showcase the flexible control over the DMA's radiation pattern, we present an experimental case study of simultaneously steering a beam to a desired transmitter and a null to an undesired jammer, achieving up to 43~dB discrimination. 
Using software-defined radios, we transmit and receive QPSK OFDM waveforms to evaluate the bit error rate.
We also discuss algorithmic and technological challenges associated with envisioned future evolutions of our end-to-end testbed and real-life DMA-based wireless systems. 
\end{abstract}

\section{Introduction}

Next-generation wireless systems require extremely flexible antennas that can tailor radiation characteristics like pattern, frequency and polarization to the needs of desired wireless functionalities. Thereby, wireless channels in dynamic environments can be substantially enhanced, unintentional or malicious interferences can be mitigated, and fine-grained sensing becomes possible. Conventional massive multiple-input multiple-output (mMIMO) systems aspire to achieving the required flexibility by deploying massive numbers of antennas, each with a dedicated radio-frequency (RF) chain. However, the practical realizability of such multi-antenna systems at large scales remains challenging because of the associated cost, power consumption, weight, and footprint. Therefore, promising alternative technologies relying on hybrid architectures are emerging, wherein reconfigurable wave-domain properties are leveraged to drastically reduce the number of required RF chains.

Various conceptually related embodiments of hybrid analog/digital antenna architectures are currently under investigation, including electrically steerable passive array radiators (ESPARs)~\cite{kawakami2005electrically}, antenna arrays connected to configurable combining circuits~\cite{gong2020rf}, dynamic metasurface antennas (DMAs)~\cite{shlezinger2021dynamic}, reconfigurable pixel antennas~\cite{ying2023reconfigurable}, and the direct integration of reconfigurable intelligent surfaces (RISs) into antenna systems~\cite{huang2025integrating}. All these architectures leverage tunable lumped elements (in most cases PIN or varactor diodes) to parametrize wave propagation within the antenna system, enabling a joint optimization of wave-domain and digital parameters that substantially reduces the required number of RF chains. Consequently, the associated cost and power consumption are expected to be drastically lower. Among the mentioned embodiments, DMAs stand out in terms of compactness because of their ultra-thin form factor. Indeed, a DMA is an ultra-thin device consisting of a surface  patterned with tunable metamaterial elements (coined meta-atoms) that are coupled to the feeds (and other meta-atoms) via flat waveguide or cavity structures. The DMA's radiation pattern is the superposition of the fields radiated by the tunable meta-atoms.

Current research on these emerging antenna technologies is still at an early stage with essentially separate thrusts on theoretical studies using simplified models and a few experimental antenna-level studies. The lack of end-to-end wireless system prototypes results in important knowledge gaps. 
\textit{First}, system-level demonstrations enforce the consideration of practical antenna-level hardware impairments (e.g., few-bit tunable lumped elements) and physically consistent models (e.g., aware of mutual coupling (MC) effects).
\textit{Second}, system-level prototyping result in confrontations with further hardware impairments beyond the antenna level, such as synchronization difficulties, phase drifts, and non-linear signal distortions in mixers and amplifiers.
\textit{Third}, system-level experiments challenge assumptions of perfectly known model parameters (e.g., antenna locations and MC strengths), triggering investigations of thus-far largely neglected MC-aware model parameter estimation techniques and/or model-agnostic optimization strategies.

Against this background, here, we present an end-to-end DMA-empowered wireless system testbed; moreover, we report a case study of jamming-resilient, frequency-agile, dynamic beam-steering evaluated in our end-to-end testbed. 

The primary goal of this paper is to inform wireless practitioners about commonly overlooked challenges that arise in end-to-end wireless system testbeds involving hybrid analog/digital antenna architectures in general, and DMAs in particular. Our article gives a flavor of the required vertical integration of expertise in antenna engineering, RF instrumentation, and signal processing. We expect that this work can, on the one hand, serve as a basis for the development of DMA-based testbeds elsewhere and, on the other hand, motivate the exploration of unveiled DMA-related system-level challenges.

We pay particular attention to examining how strong MC between the DMA's meta-atoms influences end-to-end experiments. 
The key consequence of MC is that the field radiated by any given meta-atom depends on how the other meta-atoms are configured. The mapping from the DMA's configuration to its radiation pattern is thus non-linear. MC in DMAs is so far usually mitigated and/or neglected~\cite{shlezinger2021dynamic,boyarsky2021electronically} to recover an effectively linear mapping. However, this work-around prevents wireless systems from reaping recently discovered performance benefits enabled by strong MC~\cite{prod2024mutual,prod2025benefits}. 
Our case study confronts us with the problem of optimizing a DMA with strong MC under binary programmability constraints, without disposing of an MC-aware model. Moreover, our optimization objective of simultaneous beam-and-null steering is more complex than commonly studied beam-steering~\cite{boyarsky2021electronically,DMA_glasgow,prod2025benefits}, because in particular null steering is delicate and very sensitive to model inaccuracies. These challenges lead us to explore a model-agnostic but MC-aware optimization, and to benchmark it against a model-based but MC-unaware technique. Altogether, our case study alerts wireless practitioners to important nuances in how MC-unaware optimizations perform under realistic hardware constraints and different optimization objectives.

In this article, we begin by describing our end-to-end DMA-empowered wireless system testbed, detailing the emulated scenario and the involved hardware modules. Then, we present our case study of jamming-resilient, frequency-agile, DMA-empowered, dynamic beam-steering; we detail the optimization of our DMA with strong MC, we discuss antenna-level measurements of the optimized channels, and we present our system-level performance evaluation in terms of the experimentally measured constellation diagrams and BER for different jamming strengths. We conclude by looking forward to envisioned future evolutions of our end-to-end testbed as well as algorithmic and hardware challenges in transitioning DMA-based theoretical ideas to practical deployments.

\section{End-to-End DMA Wireless System Testbed}
\label{sec_2}

\begin{figure*}[t]
\centering
\includegraphics [width = \linewidth]{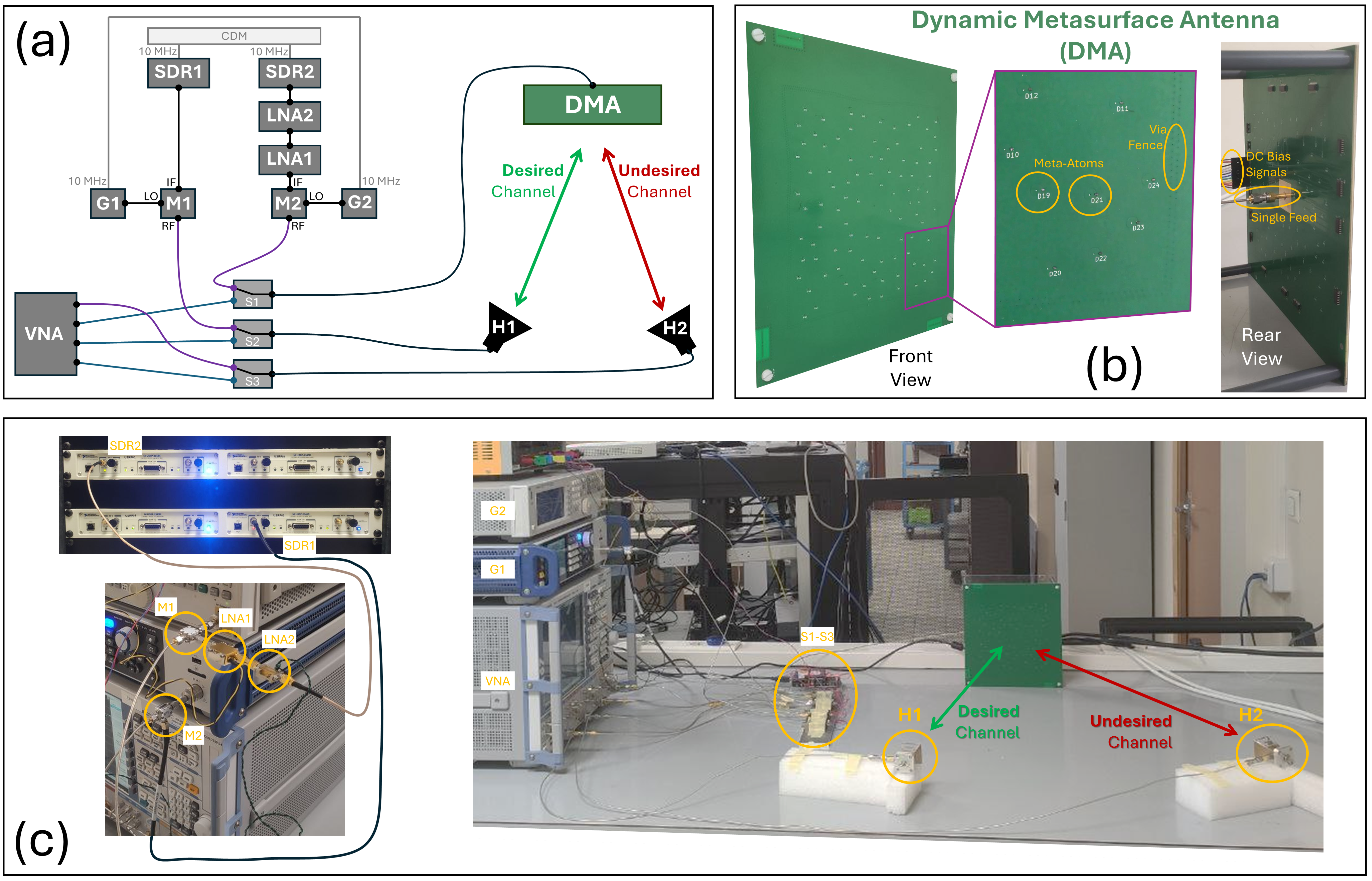}
\caption{\textbf{Hardware overview for the end-to-end DMA wireless system testbed.}  (a) Schematic of end-to-end testbed setup. Cables for power supplies and connections to the host computer are not shown. In the testbed's main mode (purple), the DMA is connected to the receiving SDR2, H1 is connected to the transmitting SDR1, and H2 is connected to a vector-network-analyzer (VNA) port that outputs the interference signal. In the auxiliary mode (blue), the three antennas (DMA, H1, H2) are connected to the three remaining VNA ports for precise end-to-end wireless channel measurements between the antenna ports. (b) Photographic images of the DMA prototype. (c) Photographic images of the testbed setup.}
\label{fig1}
\end{figure*}

In this section, we present an overview of the hardware modules of our end-to-end DMA wireless system testbed. A schematic and photographic images are shown in Fig.~\ref{fig1}. Besides its primary purpose of transmitting and receiving realistic signals (e.g., quadrature phase-shift keying (QPSK) orthogonal frequency-division multiplexing (OFDM) signals in our case study) in a DMA-based wireless system, our testbed also provides access to high-precision end-to-end channel measurements in an auxiliary mode.

Our testbed operates in the lower K-band, specifically in the vicinity of 19~GHz. This frequency range is attractive for next-generation wireless systems because it enables the realization of electrically large apertures within a limited area while avoiding the substantial path loss in higher millimeter-wave ranges. 

\subsection{Jamming-Resilient Agile Communications Scenario}
\label{subsec_scenario}

We emulate a scenario in which the information transfer from a fixed-pattern transmitter H1 to a flexible-pattern DMA-based receiver R is perturbed by (malicious or unintentional) interferences originating from another fixed-pattern transmitter H2. To facilitate jamming-resilient information transfer, the receiver configures its DMA to maximize the desired wireless channel H1-R and to \textit{simultaneously null the undesired wireless channel} H2-R. Incidentally, time reversal of this scenario is equivalent on the antenna level to physical-layer secure transmission from R to H1 because the transmission to H2 is nulled. However, the time-reversed scenario differs on the system level in terms of the required RF chains.

\subsection{DMA Prototype}

Our DMA prototype shown in Fig.~\ref{fig1}b comprises 96 binary-tunable meta-atoms that are coupled to a central coaxial feed and each other via a $15 \times 15\ \mathrm{cm}^2$ quasi-2D cavity realized on printed-circuit board (PCB). The cavity is formed by two parallel copper layers and a fence of 841 vias, and filled with a low-loss substrate. The irregular shape of the via fence breaks any potential symmetries. The meta-atoms are  placed at pseudo-random positions (subject to a minimum spacing constraint) on the upper conductor, within the perimeter of the via fence. Each meta-atom is a so-called complementary electric-LC (cELC) resonator that can be switched between two possible states by means of an integrated PIN diode. The binary control of each PIN diode's bias is provided by DC bias lines routed on lower PCB layers. We configure the DMA by sending a 96-element binary vector from Python via an Arduino microcontroller to twelve eight-bit shift registers. Further hardware details can be found in related technical papers~\cite{sleasman2020implementation,prod2025benefits}.

\subsection{RF Chains}
\label{subsec_RFchain}

\textbf{Transmission:} The transmitted baseband signal follows a simplified OFDM structure comprising 256 subcarriers (of which 152 are active data carriers), QPSK modulation at 2 bits per symbol, a 32-sample cyclic prefix, 20 OFDM symbols per frame, and a one-symbol preamble for synchronization. The transmitting SDR generates the corresponding intermediate-frequency (IF) signal with a center frequency of 2~GHz, a bandwidth of 15~MHz, and a sampling rate of 15~MS/s. This IF signal is then mixed with a continuous-wave (CW) local-oscillator (LO) signal to generate the desired RF signals.
We set the gain of the transmitting SDR to 20~dB; higher gains result in non-linear signal distortions.
The strength of the desired RF signal reaching the H1 antenna port is roughly $-24$~dBm (measured directly with a spectrum analyzer). The main reasons for the loss of signal strength upstream of H1 are the significant attenuation in the coaxial cables and the frequency up-conversion.

\textbf{Reception:} The RF chain on the receiving side resembles that on the transmitting side in reverse order, except that two low-noise amplifiers (LNAs) providing a gain of roughly 40~dB are inserted between the mixer's IF output and the receiving SDR. We set the receiving SDR's gain to 20~dB.

\textbf{Synchronization:} The two LO generators are synchronized via a 10~MHz signal to prevent drifts in their relative phases. Similarly, the two SDRs are synchronized via a 10~MHz signal. 
These synchronizations enable the identification of the start of the received frame with a basic correlation-based algorithm. We equalize the channel via zero forcing, based on a channel estimate obtained by comparing received and known transmitted symbols.

\textbf{Jamming:} We use our VNA's fourth port that is not used in our testbed's auxiliary VNA mode to generate the jamming signal. The latter is a quasi-CW signal at the center frequency of the desired transmission signal radiated by H1, and we can tune its power reaching the H2 port between roughly $-60$~dBm and $6$~dBm (measured directly with the spectrum analyzer).

\section{Case Study on Jamming-Resilient, Frequency-Agile, Dynamic Beamforming}
\label{sec_3}

In this section, we report our experimental end-to-end evaluation of the scenario outlined in Sec.~\ref{subsec_scenario}. \textit{First}, we discuss the MC-aware DMA optimization and introduce three benchmarks. \textit{Second}, we present antenna-level measurements of the optimized wireless channels in our testbed's auxiliary VNA mode. \textit{Third}, we describe the end-to-end system-level performance evaluation in our testbed's main mode.

\subsection{MC-Aware DMA Optimization and Benchmarks}
\label{subsec_MCawareOptim}

Our optimization objective is to minimize a cost defined as the undesired channel gain minus the desired channel gain, subject to the binary programmability constraint of the meta-atoms. 
A prerequisite is typically precise knowledge of the transmitter locations
-- either to inject them into a model that predicts the two channels of interest as a function of the DMA configuration to optimize the latter, or to look up a previously optimized configuration in a codebook. We assume to precisely know the transmitter locations and discuss how the DMA could sense those in Sec.~\ref{subsec_ISAC}. We choose the described codebook approach in the following because so far we do not have a calibrated MC-aware system model; we discuss how the latter may be obtained in the future in Sec.~\ref{subsec_ModelCalib}. 

To establish the codebook, we hence use a model-agnostic algorithm. For a given operating frequency and given transmitter locations, we first measure the two channels H1-R and H2-R for 500 random configurations in our testbed's auxiliary VNA mode. We select the lowest-cost configuration among the 500 considered ones as initialization for a coordinate-descent algorithm. The latter loops up to five times over each meta-atom in turn, flipping its configuration if that reduces the cost. The algorithm stops early in case of an entire loop without any updates of the configuration. 

We denote the optimized configuration by \textit{OPT}. We further consider the following  benchmarks:
\begin{itemize}
    \item \textit{MAX}: We optimize a configuration similarly to how we obtain \textit{OPT} except that we minimize an alternative cost defined as the negative of the desired-channel gain, ignoring the undesired channel.
    \item \textit{LIN}: We calibrate an MC-unaware model mapping the DMA configuration to the DMA radiation pattern using a multi-variable linear regression based on the 500 measured channel realizations. We then conduct a model-based in-software coordinate-descent optimization.
    \item \textit{RAND}: We pick a random DMA configuration.
\end{itemize}

\subsection{Antenna-Level Channel Measurements}

\begin{figure}
\centering
\includegraphics [width = \linewidth]{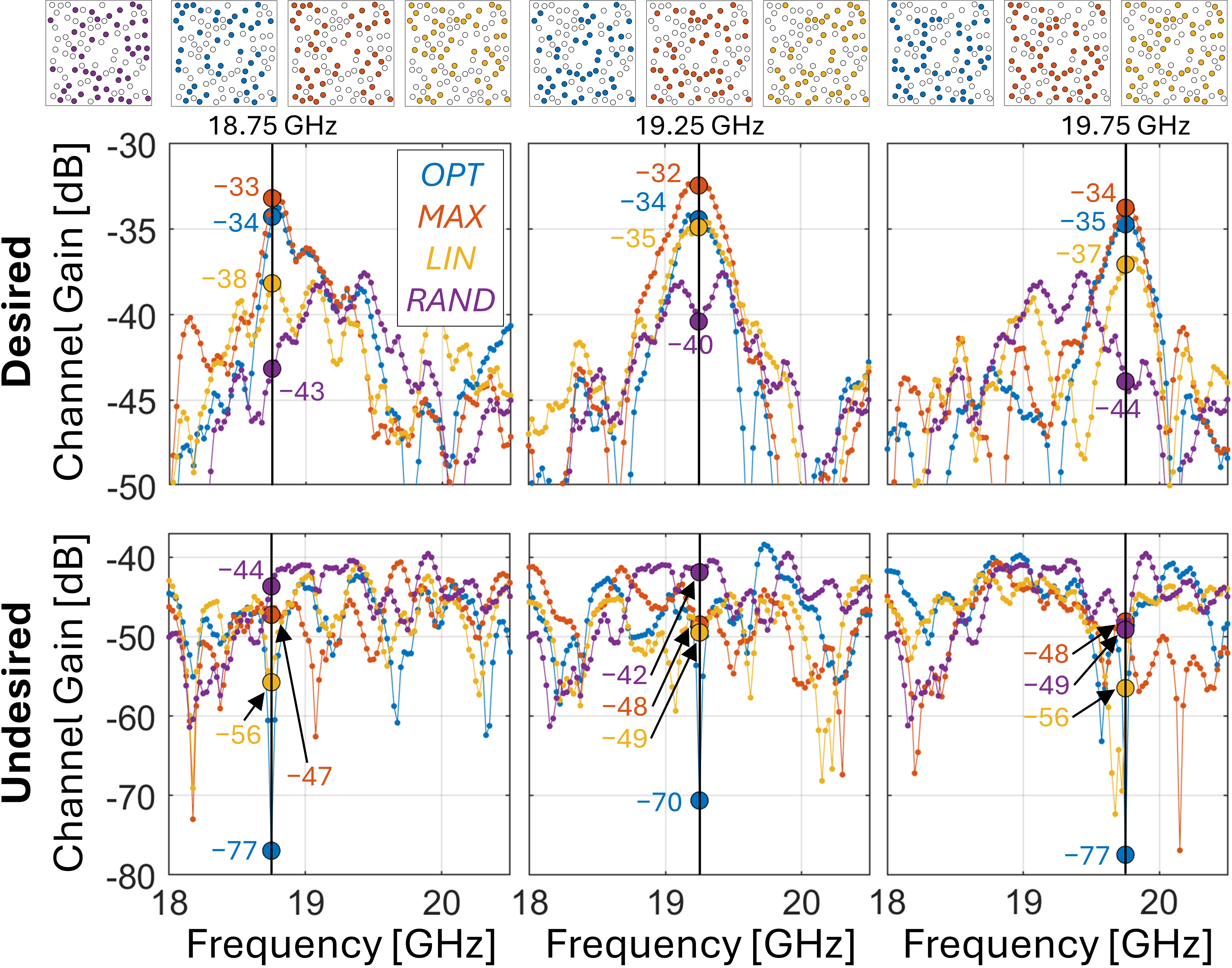}
\caption{\textbf{Measured optimized wireless channels for simultaneous beam-and-null steering.} For three operating frequencies (columns), the VNA-mode measurements of the optimized channel gains toward the desired (top row) and undesired (bottom row) transmitters are shown as a function of frequency for the four different DMA configurations displayed above (see legend). }
\label{fig2}
\end{figure}

We present in Fig.~\ref{fig2} the VNA-mode measurements of the optimized wireless channels for three selected operating frequencies (18.75~GHz, 19.25~GHz, 19.75~GHz); the transmitter locations are shown in Fig.~\ref{fig1}c. 

The \textit{OPT} DMA configuration optimized for 18.75~GHz yields a gain of $-34$~dB for the desired channel and $-77$~dB for the undesired channel, providing a strong discrimination of 43~dB between the two transmitters at the receiver. 
The spectral width of the desired-channel peak significantly exceeds the 15~MHz bandwidth of our OFDM signal. Meanwhile, the nulled undesired-channel dip is very narrow because the underlying transmission zero is a spectral singularity. A spectrally narrow null is sufficient given the narrow-band nature of our testbed's jamming signal. For more broadband jammers, alternative costs could be defined; however, there is a fundamental trade-off between the achievable depth and width of a null.
Very similar results at the other two considered operating frequencies demonstrate the DMA's frequency agility. In all optimized DMA configurations, roughly half of the meta-atoms are in either of the two possible states.

For the \textit{MAX} DMA configuration, the available radiation-pattern control is fully dedicated  to maximizing the desired-channel gain which is hence slightly higher (e.g., $-33$~dB at 18.75~GHz) than for \textit{OPT}; meanwhile, the undesired-channel gain is not at all suppressed for \textit{MAX} (e.g., $-47$~dB at 18.75~GHz), resulting in poor discrimination of only 14~dB between the two transmitters at the receiver. 

The desired-channel gain achieved with the \textit{LIN} DMA configuration based on an MC-unaware model surprisingly closely matches the one for \textit{OPT} at 19.25~GHz (only 1~dB difference) whereas it falls 4~dB short of the \textit{OPT} desired-channel gain at 18.75~GHz. It is also apparent that the \textit{LIN} and \textit{OPT} configurations are strongly correlated for all three considered operating frequencies. However, the nulling of the undesired channel with the \textit{LIN} configuration falls roughly 20~dB short of that achieved with the \textit{OPT} configurations in all cases. 
We attribute the surprisingly good performance of the MC-unaware model-based optimization in terms of maximizing the desired-channel gain to the binary programmability constraint of our DMA's meta-atoms. Nonetheless, there appears to be some realization dependence (seen here for the different operating frequencies). Meanwhile, nulling an undesired channel is a more delicate objective. The few differently programmed meta-atoms between the \textit{OPT} and \textit{LIN} configurations weakly influence the desired-channel gain maximization but strongly impact the undesired-channel gain minimization.

\subsection{Experimental System-Level Performance Evaluation}

Having optimized and measured the channels in our testbed's auxiliary VNA mode, we now switch to its main mode for an end-to-end performance evaluation by transmitting realistic QPSK OFDM signals. For each considered operating frequency and DMA configuration in turn, we transmit 167\,200 bits and estimate the BER based on the number of incorrectly received bits. Inevitably, our BER estimate's statistical uncertainty scales with both the total number of transmitted bits and the error probability.

\begin{figure}
\centering
\includegraphics [width = \linewidth]{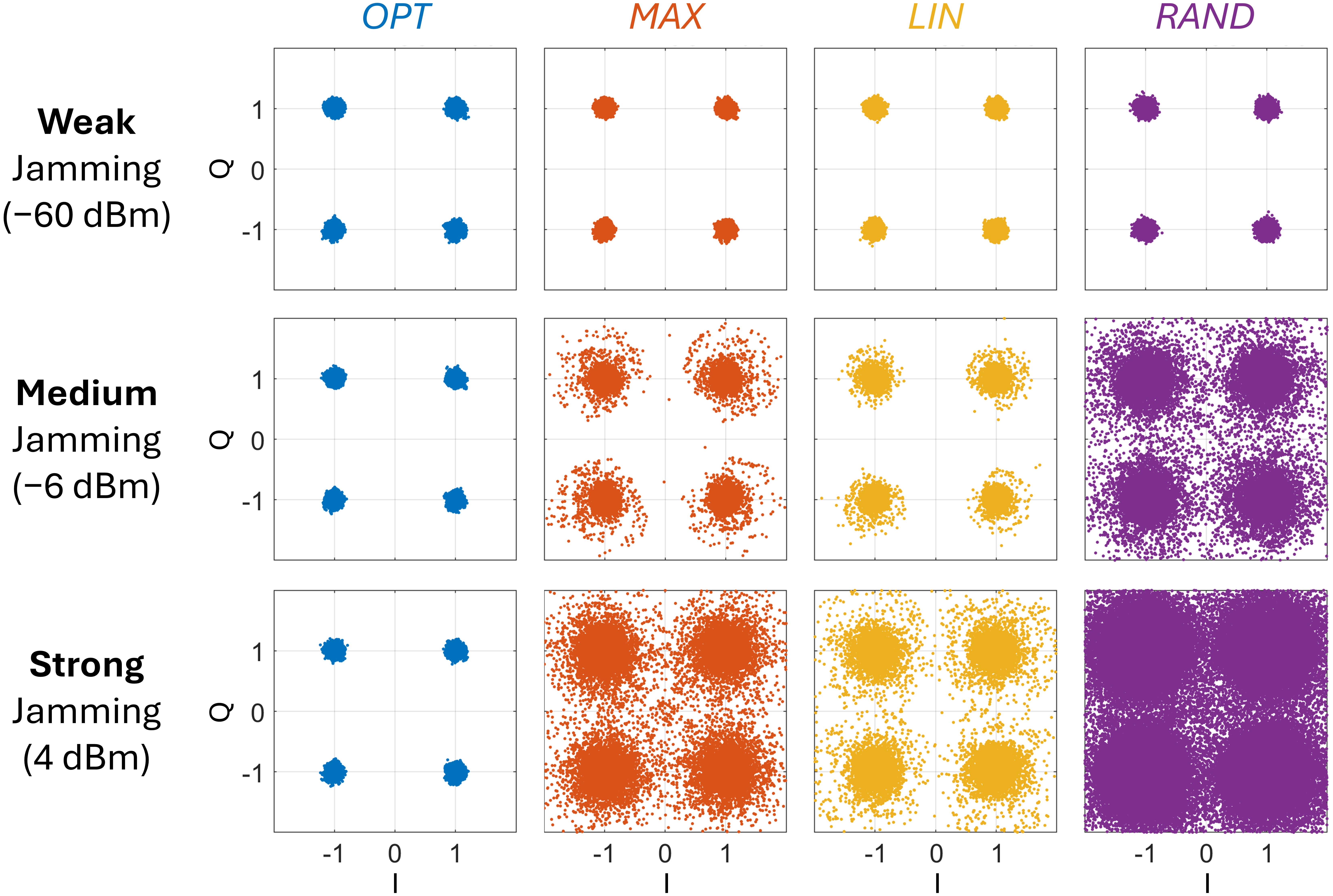}
\caption{\textbf{Measured IQ constellations for different jamming strengths and DMA configurations.} The displayed results are measured at an operating frequency of 18.75~GHz. The jamming strength denotes the power directly measured at the H2 antenna port and is indicated on the left.  }
\label{fig4}
\end{figure}

\textbf{Constellation:} We display selected IQ constellation diagrams at different jamming strengths for the four DMA configurations in Fig.~\ref{fig4}. For negligibly weak jamming, all three constellation diagrams are flawless upon visual inspection. Our chosen RF chains place us in a regime with a high signal-to-noise ratio (above 25~dB) such that the choice of DMA configuration does not notably impact the constellation in the absence of jamming. However, clear differences between the four DMA configurations appear for medium and strong jamming. Thanks to its 43~dB discrimination, the constellation of the \textit{OPT} configuration is barely affected by jamming, even when the jamming strength exceeds the desired-signal strength by 30~dB. In contrast, the deteriorating effect of jamming is clearly visible for the other three DMA configurations. \textit{RAND} is particularly vulnerable. \textit{LIN} seems to perform slightly better than \textit{MAX} because it makes an effort to achieve discrimination, albeit with limited success due to its MC-unawareness.

\begin{figure}
\centering
\includegraphics [width = 0.6\linewidth]{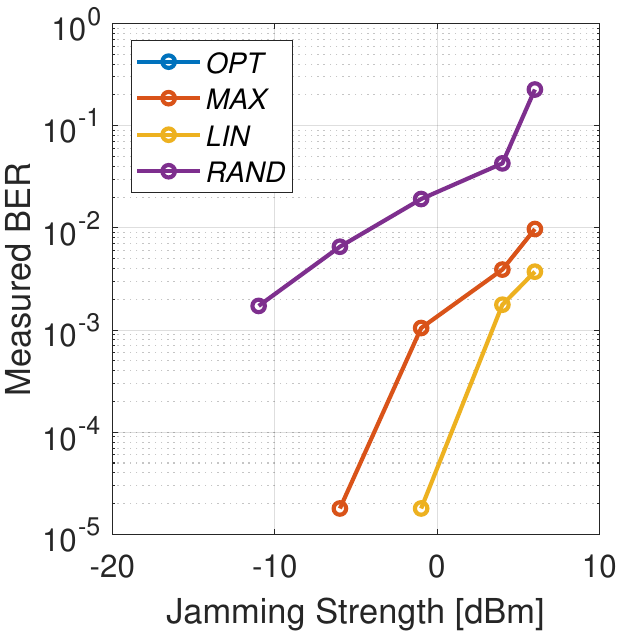}
\caption{\textbf{Measured BER as a function of the jamming strength.} The displayed results are measured at an operating frequency of 18.75 GHz. The jamming strength denotes the power directly measured at the H2 antenna port. Only data points corresponding to at least 100 incorrectly received bits out of 167\,200 transmitted bits are displayed because the statistical uncertainty of BER estimates scales with both the number of transmitted bits and the error probability. We measured zero errors for \textit{OPT} at all jamming strengths.  }
\label{fig3}
\end{figure}

\textbf{BER:} 
Representative BER estimates are displayed in Fig.~\ref{fig3} for data points corresponding to more than 100 measured errors. For weak jamming and/or for good-performing DMA configurations, certain data points are hence not displayed in Fig.~\ref{fig3}. In particular, for the \textit{OPT} configurations, none of the 167\,200 transmitted bits was incorrectly received, at all considered jamming strengths. While this does not imply a BER of zero, it implies a BER below the threshold that can be reliably estimated based on 167\,200 transmitted bits. For strong jamming with signal strengths 30~dB above the desired signal, the 43~dB discrimination at the receiver between desired and undesired channels achieved by \textit{OPT} manifestly pays off.

The \textit{LIN} configurations also seek to achieve such discrimination, but they fall substantially short of 43~dB due to MC unawareness. Consequently, while they deliver acceptable performance under weak jamming, their inferiority to the \textit{OPT} configurations obtained with MC awareness becomes very apparent under strong jamming where their BERs exceed $10^{-3}$ while \textit{OPT} still achieves a flawless performance. Similar conclusions apply to \textit{MAX} configurations for which the underlying MC-aware optimization does not attempt to null the undesired channel. The \textit{MAX} configurations are generally outperformed by the \textit{LIN} configurations in Fig.~\ref{fig3}. 
Meanwhile, the \textit{RAND} benchmark corresponds to drastically higher BERs at all jamming strengths, as expected. 
We also observe that the BER never reaches the random-guess upper bound of 0.5 in our experiments, presumably because the subcarriers outside the jammer's narrow band can always be decoded reliably.

\section{Challenges, Opportunities, \\and Research Directions}
\label{sec_challenges}

In the previous section, we provided insights into experimental end-to-end evaluations of DMA-empowered wireless systems, with particular attention to the role of MC awareness. In this section, we discuss some of the most important ensuing research challenges for wireless practitioners contributing to the realization of real-life DMA-based wireless system.

\subsection{MC-Aware vs. MC-Unaware Optimization}
\label{subsec_1}

Our case study showed that reaping the benefits of DMAs with strong MC does not always require MC-aware optimization. In particular, the combination of hardware impairments (e.g., binary-programmability constraints) with a robust optimization objective (e.g., channel gain maximization) appears to enable good performances with simple MC-unaware model-based optimizations. The gap to MC-aware optimizations may be tolerable in such cases given the drastically simpler model parameter estimation and optimization. However, the same conclusions do not apply to delicate optimization objectives (e.g., null steering), and possibly neither to DMAs with continuously tunable meta-atoms. Overall, our findings call for nuanced investigations of the circumstances under which MC-aware optimizations enable substantial performance gains that justify the associated overhead regarding model parameter estimation and optimization. Importantly, this algorithmic question should not be confused with the hardware question of whether DMA architectures with strong MC enable performance improvements~\cite{prod2024mutual,prod2025benefits}.

\subsection{MC-Aware DMA Model Calibration}
\label{subsec_ModelCalib}

A known, accurate system model is a prerequisite for many wireless communications  algorithms. For DMAs with strong MC, this requires parameter-estimation techniques for MC-aware models, as least for certain combinations of hardware constraints and optimization objectives (see Sec.~\ref{subsec_1}). 
While MC-aware DMA models exist, how to estimate their parameters for an experimentally given DMA prototype remains an open challenge. One can generally not rely on numerical simulations for parameter estimation because of fabrication inaccuracies and component tolerances.
Related experimentally validated parameter estimation techniques for RIS-parametrized channels~\cite{sol2024experimentally} can presumably be extended to DMAs.

A calibrated, accurate, MC-aware DMA model, combined with precise knowledge of the transmitter locations (see Sec.~\ref{subsec_ISAC}), will remove the reliance on our testbed's VNA mode (other than for ground-truth channel measurements). Indeed, a calibrated model can predict the channels of interest, as a function of the DMA configuration and the transmitter locations.

\subsection{Integrated Sensing and Communications (ISAC)}
\label{subsec_ISAC}

Our case study revealed the importance of sensing the transmitter locations which are required to optimize the DMA-based wireless communications system. A convergence of sensing and communications is hence pivotal~\cite{holo_isac}, and not merely motivated by limited resources. The sensing capabilities of DMA-empowered wireless systems can extend substantially beyond localization, as hinted at in specialized studies on DMA-based computational imaging~\cite{sleasman2020implementation,del2020learned}. Ultimately, the DMA-based wireless system may be capable of acquiring detailed context awareness by localizing users and obstacles, recognizing postures, etc. This context information can be injected into the system model to optimize the DMA for wireless communications.

Existing experiments on DMA-based computational imaging like~\cite{sleasman2020implementation} rely on model-agnostic dictionaries containing random DMA configurations and corresponding radiation patterns, obtained via cumbersome near-field scans. The availability of a calibrated, accurate DMA model (see Sec.~\ref{subsec_ModelCalib}) will obviate such dictionaries, and furthermore enable more advanced joint optimizations of the DMA configuration and the digital layers (see Sec.~\ref{subsec_E2E}).

Scheduling the co-existence of communications and sensing in DMA-empowered wireless systems will require a delicate balance. 
A simple solution consists in orthogonal time sharing~\cite{huang2025integrating}, i.e., sensing and communicating in separate time slots with optimized allocation. 
To minimize interruptions of wireless communications, at least some sensing may be conducted based on the signals captured in the communications mode. 
If the captured signals were originally emitted by the DMA, the sensing is more self-reliant but also more likely to temporarily block wireless communications.

\subsection{Joint Physical and Digital Layer Optimization}
\label{subsec_E2E}

To optimally reap the wireless system's DMA-empowered flexibility, all adjustable parameters (both in the wave domain and the digital layer) should ideally be jointly optimized. Theoretical explorations of task-specific joint optimizations exist in areas like sensing~\cite{del2020learned} and MIMO-OFDM communications~\cite{wang2020dynamic}. Similar concepts can be envisioned for jamming-resilient communications, such as jointly adjusting the DMA configuration and the digital post-processing to a given jammer. Ideally, joint physical-digital optimizations account for the task as well as noise, interferences, and analog-to-digital conversions. 
A common prerequisite is  an accurate, differentiable system model, emphasizing once more the importance of the MC-aware DMA model calibration discussed in Sec.~\ref{subsec_ModelCalib}. Ultimately, the joint optimization of physical and digital parameters across multiple functionalities can be envisioned for ISAC (see Sec.~\ref{subsec_ISAC}). However, it remains unknown whether the promising results on joint physical-digital optimization from theoretical studies like~\cite{del2020learned,wang2020dynamic} can be confirmed in an experimental end-to-end wireless system. Moreover, it is so far unclear whether the practically achievable benefits justify the required overhead.

\subsection{Beyond-Diagonal DMA (BD-DMA) Hardware}
\label{subsec_BDDMA}

Because hybrid/analog antenna architectures aim to program wave-domain properties to alleviate the number of required RF chains, it is important to maximize the wave-domain programmability. Conventional DMAs only offer local wave-domain programmability in terms of the meta-atoms' configuration. Meanwhile, the intrinsic MC between the feeds and meta-atoms is fixed by the cavity or waveguide structure. To break this limitation and achieve a maximally flexibile hybrid analog/digital DMA-based wireless system, BD-DMAs with \textit{tunable} intrinsic MC between their meta-atoms are enticing~\cite{prod2025beyond}. The benefits of tunable intrinsic MC grow with the DMA's MC strength~\cite{prod2025beyond}. 
Important future research directions thus concerns practical realizations of BD-DMAs and matching algorithmic solutions. Based on the physics-compliant diagonal representation of BD-DMAs, physics-compliant algorithmic solutions originally developed for conventional DMAs can be directly applied to BD-DMAs~\cite{prod2025beyond}. Moreover, similar to our case study, BD-DMA prototypes can be evaluated end-to-end with model-agnostic techniques.

\section{Conclusion}

DMAs are an emerging ultra-thin hybrid analog/digital reconfigurable antenna technology  poised to play an important role in building energy-efficient, low-cost, and compact next-generation wireless systems. In this article, we addressed the current lack of experimental end-to-end DMA-empowered wireless system testbeds that results in important knowledge gaps. 
Our described confrontations with hardware impairments on the antenna and system levels, as well as challenges due to unknown model parameters, identify important future research directions for wireless practitioners. In particular, the system-level influence of strong MC in DMAs requires a nuanced analysis specific to the hardware constraints and optimization objectives.
Our case study emulated a complex scenario of jamming-resilient, agile, DMA-aided communications. For the first time, we experimentally showcased how single-feed DMA hardware can achieve very strong discrimination between desired and undesired channels, provided that MC is accounted for in the optimization. Our experimental end-to-end  evaluation confirmed substantial performance benefits in terms of constellation diagrams and BERs.
We also discussed open research questions on realizating DMA-empowered wireless systems and rigorously evaluating emerging DMA-aided concepts that have so far remained theoretical.

\bibliographystyle{IEEEtran}
%\bibliography{references}

\begin{thebibliography}{10}
\providecommand{\url}[1]{#1}
\csname url@samestyle\endcsname
\providecommand{\newblock}{\relax}
\providecommand{\bibinfo}[2]{#2}
\providecommand{\BIBentrySTDinterwordspacing}{\spaceskip=0pt\relax}
\providecommand{\BIBentryALTinterwordstretchfactor}{4}
\providecommand{\BIBentryALTinterwordspacing}{\spaceskip=\fontdimen2\font plus
\BIBentryALTinterwordstretchfactor\fontdimen3\font minus
  \fontdimen4\font\relax}
\providecommand{\BIBforeignlanguage}[2]{{%
\expandafter\ifx\csname l@#1\endcsname\relax
\typeout{** WARNING: IEEEtran.bst: No hyphenation pattern has been}%
\typeout{** loaded for the language `#1'. Using the pattern for}%
\typeout{** the default language instead.}%
\else
\language=\csname l@#1\endcsname
\fi
#2}}
\providecommand{\BIBdecl}{\relax}
\BIBdecl

\bibitem{kawakami2005electrically}
H.~Kawakami and T.~Ohira, ``Electrically steerable passive array radiator
  ({ESPAR}) antennas,'' \emph{IEEE Antennas Propag. Mag.}, vol.~47, no.~2, pp.
  43--50, 2005.

\bibitem{gong2020rf}
T.~Gong, N.~Shlezinger, S.~S. Ioushua, M.~Namer, Z.~Yang, and Y.~C. Eldar,
  ``{RF} chain reduction for {MIMO} systems: A hardware prototype,'' \emph{IEEE
  Syst. J.}, vol.~14, no.~4, pp. 5296--5307, 2020.

\bibitem{shlezinger2021dynamic}
N.~Shlezinger, G.~C. Alexandropoulos, M.~F. Imani, Y.~C. Eldar, and D.~R.
  Smith, ``Dynamic metasurface antennas for {6G} extreme massive {MIMO}
  communications,'' \emph{IEEE Wirel. Commun.}, vol.~28, no.~2, pp. 106--113,
  2021.

\bibitem{ying2023reconfigurable}
K.~Ying, Z.~Gao, S.~Chen, X.~Gao, M.~Matthaiou, R.~Zhang, and R.~Schober,
  ``Reconfigurable massive {MIMO}: Harnessing the power of the electromagnetic
  domain for enhanced information transfer,'' \emph{IEEE Wirel. Commun.},
  vol.~31, no.~3, pp. 125--132, 2024.

\bibitem{huang2025integrating}
Y.~Huang, L.~Zhu, and R.~Zhang, ``Integrating base station with intelligent
  surface for {6G} wireless networks: Architectures, design issues, and future
  directions,'' \emph{IEEE Wirel. Commun.}, 2025.

\bibitem{boyarsky2021electronically}
M.~Boyarsky, T.~Sleasman, M.~F. Imani, J.~N. Gollub, and D.~R. Smith,
  ``Electronically steered metasurface antenna,'' \emph{Sci. Rep.}, vol.~11,
  no.~1, p. 4693, 2021.

\bibitem{prod2024mutual}
H.~Prod'homme and P.~del Hougne, ``Mutual coupling in dynamic metasurface
  antennas: Foe, but also friend,'' \emph{arXiv:2412.01002, IEEE Wirel. Commun.
  (in press)}, 2024.

\bibitem{prod2025benefits}
H.~Prod'homme, J.~Tapie, L.~Le~Magoarou, and P.~del Hougne, ``Benefits of
  mutual coupling in dynamic metasurface antennas for optimizing wireless
  communications--theory and experimental validation,''
  \emph{arXiv:2502.15565}, 2025.

\bibitem{DMA_glasgow}
A.~Jabbar, M.~Elsayed, J.~Ur~Rehman~Kazim, Z.~Pang, J.~Le~Kernec, M.~Ali~Imran,
  Q.~H. Abbasi, and M.~Ur-Rehman, ``60 {GHz} programmable dynamic metasurface
  antenna ({DMA}) for next-generation communication, sensing, and imaging
  applications: From concept to prototype,'' \emph{IEEE Open J. Antennas
  Propag.}, vol.~5, no.~3, pp. 705--726, 2024.

\bibitem{sleasman2020implementation}
T.~A. Sleasman, M.~F. Imani, A.~V. Diebold, M.~Boyarsky, K.~P. Trofatter, and
  D.~R. Smith, ``Implementation and characterization of a two-dimensional
  printed circuit dynamic metasurface aperture for computational microwave
  imaging,'' \emph{IEEE Trans. Antennas Propag.}, vol.~69, no.~4, pp.
  2151--2164, 2020.

\bibitem{sol2024experimentally}
J.~Sol, H.~Prod'homme, L.~Le~Magoarou, and P.~del Hougne, ``Experimentally
  realized physical-model-based frugal wave control in metasurface-programmable
  complex media,'' \emph{Nat. Commun.}, vol.~15, no.~1, p. 2841, 2024.

\bibitem{holo_isac}
H.~Zhang, H.~Zhang, B.~Di, M.~Di~Renzo, Z.~Han, H.~V. Poor, and L.~Song,
  ``Holographic integrated sensing and communication,'' \emph{IEEE J. Sel.
  Areas Commun.}, vol.~40, no.~7, pp. 2114--2130, 2022.

\bibitem{del2020learned}
P.~del Hougne, M.~F. Imani, A.~V. Diebold, R.~Horstmeyer, and D.~R. Smith,
  ``Learned integrated sensing pipeline: reconfigurable metasurface
  transceivers as trainable physical layer in an artificial neural network,''
  \emph{Adv. Sci.}, vol.~7, no.~3, p. 1901913, 2020.

\bibitem{wang2020dynamic}
H.~Wang, N.~Shlezinger, Y.~C. Eldar, S.~Jin, M.~F. Imani, I.~Yoo, and D.~R.
  Smith, ``Dynamic metasurface antennas for {MIMO-OFDM} receivers with
  bit-limited {ADCs},'' \emph{IEEE Trans. Commun.}, vol.~69, no.~4, pp.
  2643--2659, 2020.

\bibitem{prod2025beyond}
H.~Prod'homme and P.~del Hougne, ``Beyond-diagonal dynamic metasurface
  antenna,'' \emph{arXiv:2504.13523}, 2025.

\end{thebibliography}

% Generated by IEEEtran.bst, version: 1.14 (2015/08/26)

\end{document}